\documentclass{elsart}

\usepackage{fleqn}
\usepackage{graphicx}
\usepackage{epsfig}
\usepackage{amsmath}
\usepackage{amsfonts}
\usepackage{amssymb}

\def\be{\begin{equation}}
\def\ee{\end{equation}}
\def\bea{\begin{eqnarray}}
\def\eea{\end{eqnarray}}
\def\ep{\epsilon}

\def\T{{\mathcal F}}
\def\N{{\mathcal G}}
\def\TT{{\mathcal T}}
\def\NN{{\mathcal V}}
\def\t{\tau_z}
\begin{document}

\begin{frontmatter}

\title{\bf Simple solutions of relativistic hydrodynamics\\
	for longitudinally and cylindrically expanding systems}

\author[KFKI,USP]{T. Cs\"org\H o,\thanksref{tamas}}
\author[USP]{F. Grassi,\thanksref{frederique}}
\author[USP]{Y. Hama,\thanksref{yogiro}}
\author[UFRJ]{T.  Kodama\thanksref{takeshi}}
\address[KFKI]{MTA KFKI RMKI, H - 1525 Budapest 114, POBox 49, Hungary}
\address[USP]{IF-USP, C.P. 66318, 05389-970 S\~{a}o Paulo, SP, Brazil}
\address[UFRJ]{IF-UFRJ, C.P. 68528, 21945-970 Rio de Janeiro, RJ, Brazil}
\thanks[tamas]{Email: csorgo@sunserv.kfki.hu}
\thanks[frederique]{\phantom{Email:} grassi@if.usp.br}
\thanks[yogiro]{\phantom{Email:} hama@fma.if.usp.br}
\thanks[takeshi]{\phantom{Email:} tkodama@if.ufrj.br}

\date{Jan. 21, 2003} 
\begin{abstract}
Simple, self-similar, analytic solutions of 1 + 1 dimensional relativistic
hydrodynamics are presented, generalizing the Hwa - Bjorken
boost-invariant solution to inhomogeneous
rapidity distributions. These solutions are generalized also to 
1 + 3 dimensional, cylindrically symmetric firetubes, 
corresponding to central collisions of heavy ions at 
relativistic bombarding energies.
\end{abstract}

\end{frontmatter}

\date{February 3, 2003}
\maketitle

\section{Introduction}
Analytic solution of the equations of relativistic hydrodynamics 
is a difficult task because the equations are non-linear 
partial differential equations, that are rather complicated to handle
not only analytically but also numerically. 
However, relativistic hydrodynamics has various applications, including the
calculations of single-particle spectra and two-particle correlations in 
relativistic heavy ion collisions, see ref~\cite{csernai}. More recently, there has been an increasing 
interest in applications of relativistic hydrodynamics 
in Au+Au collisions at  RHIC both at $\sqrt{s}=130$ AGeV 
and $\sqrt{s}=200$AGeV bombarding
energies, predictions were made for the coming LHC experiments 
\cite{Shu,Hirano,Kolb}. 
The hydrodynamical analysis can also be extended to the study of these 
processes on event-by-event basis \cite{SPH-jpg,SPH-qm01}. However, 
most works in hydrodynamics are numerical so not always transparent. 

In this sense, exact solutions would be useful, but are rarely found 
due to the highly non-linear nature of relativistic hydrodynamics. Khalatnikov's one-dimensional analytical solution \cite{Khalatnikov} 
to Landau's hydrodynamic model \cite{Landau} gave rise to a new 
approach in high energy physics. The boost-invariant solution 
\cite{scale-invariant} was found later by R. C. Hwa
and other authors. It has been frequently utilized 
as the basis for estimations of initial energy densities in 
ultra-relativistic nucleus-nucleus collisions\cite{Bjorken}. 
Due to this famous application this boost-invariant
solution is frequently called as Bjorken's solution,
although as far as we know it was first described by R. Hwa 
in ref.~\cite{scale-invariant}. Perhaps it should be called
the Hwa-Bjorken solution, which name we shall use hereafter. 

Recently, 
Bir\'{o} has found  self-similar exact solutions of relativistic 
hydrodynamics for cylindrically expanding systems \cite{biro1,biro2}. 
However, his solutions are valid only when the pressure is independent 
of space and time, as e.g. in the case of a rehadronization phase 
transition in the middle of a relativistic heavy ion collision.

Here we present an analytic approach, which goes back to the 
data-motivated exact analytic solution of non-relativistic 
hydrodynamics found by Zim\'{a}nyi, Bondorf and Garpman (ZBG) in 1978 
for low energy heavy ion collisions with spherical symmetry~\cite{jnr}. 
This solution has been extended to the case of elliptic symmetry 
by Zim\'{a}nyi and collaborators in ref.~\cite{jde}. 
In ~\cite{nr,nrt} a Gaussian parameterization has been introduced to
describe the mass dependence of the effective temperature and the 
radius parameters of the two-particle Bose-Einstein correlation 
functions in high energy heavy ion collisions. Later it has been 
realized that this phenomenological \textit{parameterization} of data 
corresponds to an exact, Gaussian \textit{solution} of non-relativistic 
hydrodynamics with spherical symmetry~\cite{cspeter}. The spherically 
symmetric self-similar solutions of non-relativistic hydrodynamics were 
obtained in a general manner in \cite{cssol}, that included an arbitrary 
scaling function for the temperature profile, and expressed the density 
distribution in terms of the temperature profile function. The ZBG 
solution and the Gaussian solution of \cite{cspeter} are recovered from 
the general solution of ~\cite{cssol} as special cases, corresponding to 
different scaling functions of the temperature profile. The Gaussian 
solution has been generalized to ellipsoidal expansions in \cite{ellsol}, 
that provides analytic insight into the physics of non-central heavy ion 
collisions~\cite{ellsp}.

Our approach corresponds to a generalization of these recently 
obtained analytic solutions~\cite{cspeter,cssol,ellsp,csell} 
of non-relativistic fireball hydrodynamics to the case of 
relativistic longitudinal and transverse flows. In particular, an 
analytic approach, the Buda-Lund (BL) model has been developed to 
{\it parameterize} the single particle spectra and the two-particle 
Bose-Einstein correlations in high-energy heavy-ion physics in 
terms of hydrodynamically expanding, cylindrically symmetric 
sources~\cite{3d}. Here we attempt to find a {\it family of exact 
solutions} of relativistic hydrodynamics that may include the BL 
model as a particular limiting case.
It turns out that in the simplest case our result corresponds to
the Cracow hydrodynamic parametization, 
which is successfull in describing single particle spectra
of Au+Au collisions at $\sqrt{s}=130$ and 200 AGeV at RHIC\cite{cracow,cracow-rev,cracow-school}.

\section{The equations of relativistic hydrodynamics} 
We solve the relativistic continuity and energy-momentum conservation
equation: 
\begin{eqnarray}
\partial_\mu(n u^\mu) & = & 0\,, \label{e:cont} \\
\partial_\nu T^{\mu \nu} & = & 0\,.  \label{e:tnm} 
\end{eqnarray}
Here $n \equiv n(t,{\mathbf r})$ is the number density, the 
four-velocity is denoted by 
$u^\mu \equiv u^\mu(t,{\mathbf r}) = \gamma (1, {\mathbf v})$, 
normalized to $u^\mu u_\mu = \gamma^2 (1 - {\mathbf v}^2) = 1$, and 
the energy-momentum tensor is denoted by $T^{\mu \nu}$. We assume 
perfect fluid, 
\begin{equation} 
T^{\mu\nu} = (\ep + p) u^\mu u^\nu - p g^{\mu \nu},
\end{equation} 
where $\ep$ stands for the relativistic energy density and $p$ denotes the pressure.

We close this set of relativistic hydrodynamical equations 
with the equations of state. We assume a gas containing massive 
conserved quanta, 
\bea
\ep & =& m n + \kappa p\,, \label{e:eos1}\\
p  & = & n T \label{e:eos2}.
\eea
The equations of state have two free parameters, $m$ and $\kappa$.
Non-relativistic hydrodynamics of ideal gases corresponds to the
limiting case  of 
$m \gg T$, ${\mathbf v}^2 \ll 1$ and $\kappa = 3/2\,$. Relativistic
hydrodynamics for massless particles and a constant speed of sound
$c_s^2$ corresponds to the case of $m=0$ and $c_s^2 = 1/\kappa$.

The energy-momentum conservation equations can be projected into a 
component parallel to $u^\mu$ and components orthogonal to $u^\mu$, 
which are respectively the relativistic energy and Euler equations: 
\bea
u^\mu \partial_\mu \ep + (\ep + p) \partial_\mu u^\mu & = & 0\,, \label{e:ren}\\
u_\nu u^\mu \partial_\mu p + (\ep + p)u^\mu \partial_\mu u_\nu - 
\partial_\nu p & = & 0\,. \label{e:rEu}
\eea 
Based on general thermodynamical considerations, one can show that
the expansion is adiabatic:
\be
\partial_\mu(\sigma u^\mu)  =  0\,, \label{e:scont}
\ee
where $\sigma$ is the entrophy density. 
This relation holds for perfect fluids,
 independently of the equations of state. 

With the help of the equations of state and the continuity equation, 
the energy equation  can be rewritten as an equation for the 
temperature, 
\begin{equation} 
u^\mu \partial_\mu T + \frac{1}{\kappa} T \partial_\mu u^\mu = 0\,. \label{e:rT}
\end{equation} 

We solve 5 independent equations, the continuity, 
the (3 spatial  components of) relativistic Euler, 
and the temperature equation, 
eqs.~(\ref{e:cont},\ref{e:rEu},\ref{e:rT}).
The equations of state, eq.~(\ref{e:eos1},\ref{e:eos2}) close this system of equations in terms of 5 variables, $n$, $T$ and 
${\mathbf v} = (v_x,v_y,v_z)$. 

\section{Self-similarity } 
We look for solutions which generalize the usual similarity flow, in which 
the flow pattern is unchanged with time if the scales of length 
$X(t), Y(t), Z(t)$ along three orthogonal directions vary appropriately, 
namely, we consider 
\begin{equation}
{\mathbf v} = \left(\frac{\dot X(t)}{X(t)}r_x, 
                    \frac{\dot Y(t)}{Y(t)}r_y, 
                    \frac{\dot Z(t)}{Z(t)}r_z\right)\,, 
\end{equation}
where $x^\mu\equiv(t,r_x\,,r_y\,,r_z)$ and the dot indicates the time 
derivative. As for the thermodynamic quantities such as $n(x^\mu), 
T(x^\mu), p(x^\mu), \ldots,$ we search solutions of the form 
\begin{equation} 
f(x^\mu) = f_0 \left(\frac{V_0}{V}\right)^a F(s)\,,
\end{equation}
where the volume parameter $V=XYZ$, $a$ is an appropriate exponent and 
$F(s)$ is an arbitrary fuction of the scaling variable defined by 
\begin{equation} 
s=\frac{r_x^2}{X^2} +\frac{r_y^2}{Y^2} + \frac{r_z^2}{Z^2}\ . \label{e:sdef} 
\end{equation} 

These are Hubble type of flows, but the thermodynamic quantities may contain 
arbitrary functions depending on the the scale parameter $s$ and also, at 
least in principle, the scale parameters $X(t),Y(t)$ and $Z(t)$ may 
be different in the principal directions. Their derivatives,
$\dot X(t)$, $\dot Y(t)$ and $\dot Z(t)$ correspond to (direction and time
dependent, generalized) Hubble constants.

In heavy-ion collisions, the well known boost-invariant solution 
\cite{scale-invariant} is often utilized to discuss several properties of 
data. However, this solution has some shortcomings: {\it i)} it is scale 
invariant, having a flat rapidity distribution, corresponding to the 
extreme relativistic collisions; {\it ii)} it contains no transverse flow. 
In the present paper, we apply the strategy described above first to 1+1 
dimensional (time + longitudinal coordinate) case and obtain a class of 
solutions which are able to describe inhomogeneous rapidity distributions, 
overcoming the first shortcoming mentioned above. Then, in section \ref{s:cyl}, we 
consider the case of cylindrically symmetric case, trying to overcome the 
second shortcoming.

\section{Simple 1+1 dimensional solutions}

In this section, we solve the 1+1 dimensional problem. Hence 
$x^\mu = (t,r_z)$, $k^\mu = (E, k_z)$ throughout this section. The metric 
tensor is $g^{\mu\nu} = g_{\mu\nu} = \mbox{\rm diag}(1,-1)$ and 
$x_\mu = (t, -r_z)$.
We solve 3 independent equations, the continuity, the temperature equation 
and the $z$ component of the Euler equation 
(\ref{e:cont},\ref{e:rEu},\ref{e:rT}). 
The equations, (\ref{e:eos1},\ref{e:eos2}) close this system of equations 
in terms of 3 variables, $n$, $T$ and $v_{z}$. 

We look for flows that scale in the $z$ direction. The scaling variable, 
eq.(\ref{e:sdef}), in this case is defined as 
\begin{equation} 
s=\frac{r_{z}^{2}}{Z\left(  t\right)  ^{2}},
\end{equation}
and the longitudinal velocity 
\begin{equation}
v_{z}(t,r_{z})=\frac{\dot{Z}(t)}{Z(t)}r_{z},\label{e:vz1}%
\end{equation}
where $\dot{Z}=dZ(t)/dt\,$. In the relativistic notation, this form is 
equivalent to 
\begin{eqnarray} 
u^{\mu}  & = & (\cosh\zeta,\sinh\zeta),\\
\tanh\zeta & = & \frac{\dot{Z}(t)}{Z(t)}r_{z}, \qquad \mbox{\rm or} \quad
\cosh\zeta \, = \, \frac{1}{\sqrt{1-\dot{Z}^{2}s}}\equiv\gamma.\label{e:zeta}
\end{eqnarray}
Note that from eq. (\ref{e:zeta}) it is obvious that this solution
can be defined only in a bounded longitudinal coordinate region, because
at any time $|r_z| \le  Z(t)/\dot Z(t)$ has to be satisfied.
Using this ansatz, we find that the continuity equation is solved by the form
\begin{equation}
n(t,r_{z})=n_{0}\frac{Z_{0}}{Z}\frac{1}{\cosh\zeta}
\mathcal{G}(s),\label{s:cont1} 
\end{equation}
where $\mathcal{G}(s)$ is an arbitrary non-negative function of the scaling
variable $s$ and $n_{0}$ and $Z_{0}$ are normalization constants. We use the
convention $Z_{0}=Z(t_{0})$ and $n_{0}=n(t_{0},0)$ which implies that
$\mathcal{G}(s=0)=1$. The temperature equation, (\ref{e:rT}) is solved by the
following form:
\begin{equation}
T(t,r_{z})=T_{0}\left(  \frac{Z_{0}}{Z}\frac{1}{\cosh\zeta}\right)
^{1/\kappa}\mathcal{F}(s).
\end{equation}
The constants of normalization are chosen such that $T_{0}=T(t_{0},0)$ and
$\mathcal{F}(0)=1$. Here again, we find that the solution is independent of
the form of the function $\mathcal{F}(s)$.
From the positivity of the temperature distribution it follows that 
$\mathcal{F}(s)\geq0$.

Using the ansatz for the flow profile and the solution for the density and the
temperature, the relativistic Euler equation reduces to a complicated
non-linear equation that contains $Z$, $\dot{Z}$ and $\ddot{Z}$ and $s$.
Taking this equation at $s=0$ we express $\ddot{Z}$ as a function of $Z$ and
$\dot{Z}$. Substituting this back to the Euler equation we obtain an equation
for $\dot{Z},Z$ and $s$. In particular, for the $m=0$ case, $Z$ cancels out
and this reduces to a second order polynomial equation for $\dot{Z}^{2}$,
which has only one positive root. The form of the solution in this case
($m=0$) is
$\dot{Z}^{2}(t)=F(s)$.
Observing that the function $F$ depends only on the scaling variable $s$,
while $\dot{Z}$ depends only on the time variable $t$, we conclude that the
only solution of this equation should be a constant
$\dot{Z}=\dot{Z}_{0}$.
Now we choose the origin of the time axis such that $Z(t=0)=0$ without loss of
generality. The solutions can be cast in a relatively simple form by
introducing the longitudinal proper time $\tau$ and the space-time rapidity
$\eta$, 
\begin{eqnarray}
\tau & = & \sqrt{t^{2}-r_{z}^{2}},\label{e:tau1}\\
\eta & = & 
	\frac{1}{2}\log\left(  \frac{t+r_{z}}{t-r_{z}}\right)  .\label{e:eta1}%
\end{eqnarray}
This implies that
$Z(t)   =  \dot{Z}_{0}t$, $v_{z}  = \frac{r_{z}}{t} = \tanh\eta$ and $\zeta  =  \eta$.
Thus the solution for the flow velocity field corresponds to the flow field 
of the boost-invariant solution. However, in the boost-invariant solution the 
temperature distribution was independent of the $\eta$ variable, while in 
our case the density and the temperature distributions can be both $\eta$ 
dependent, or in other words, our solutions are scale dependent. The scale 
is defined by the parameter $\dot{Z}_{0}$, in the longitudinal direction. 

This special form of the solution for the flow velocity field implies that
$\ddot Z = 0$. This equation implies that there is no pressure gradient and
there is no acceleration in this class of self-similar solutions, 
similarly to the case of boost-invariant solution. The Euler equation is 
reduced to the following requirement:
\begin{equation}
(\partial_{z} + \frac{r_{z}}{t} \partial_{t}) \left[  \left( \frac{ t_{0}
}{\tau}\right) ^{(1+1/ \kappa)} ( 1 - \dot Z_{0}^{2} s)^{(1+1/\kappa)}
\mathcal{G}(s) \mathcal{F}(s) \right]  = 0
\end{equation}
This equation is solved by the trivial $\mathcal{G}(s) \mathcal{F}(s) = 0$ as
well as by the non-trivial solution of
\begin{equation}
\mathcal{G}(s)\mathcal{F}(s) = (1 - \dot Z_0^2 s)^{-(1 + 1/\kappa)}, 
\label{e:nt1} 
\end{equation}
which is indeed only a function of $s$ as $\dot Z_{0}$ is a constant of time.
With this form, the Euler equation is satisfied. This solution implies that
the scaling profile functions for the temperature and the density distribution
are not independent. As the constraint is given only for their product, one of
them can be still chosen in an arbitrary manner.

It is worthwhile to introduce new forms of the scaling
functions. Let us define
\begin{eqnarray}
\mathcal{T}(s)  & = & \mathcal{F}(s)(1-\dot{Z}_{0}^{2}s)^{1/\kappa},  \\
\mathcal{V}(s)  & = & \mathcal{G}(s)(1-\dot{Z}_{0}^{2}s)\,.  
\end{eqnarray}
Then the constraint Eq.
(\ref{e:nt1}) can be cast to the simplest form of
\begin{equation}
\mathcal{V}(s)\mathcal{T}(s)=1.
\end{equation}
Let us summarize our new family of solutions of the 1+1 dimensional
relativistic hydrodynamics by substituting the results in the density,
temperature and pressure profiles.
We obtain
\begin{eqnarray}
v_{z}  & = &\frac{r_{z}}{t}=\tanh\eta,\\
s      & = & 
	\frac{r_{z}^{2}}{\dot{Z}_{0}^{2}t^{2}}
	\,=\,\frac{\tanh^{2}\eta}{\dot {Z}_{0}^{2}},\\
n  & = & n_{0}\frac{t_{0}}{\tau}\mathcal{V}(s),\label{e:nsol1}\\
p  & = & p_{0}\left(  \frac{t_{0}}{\tau}\right)  ^{1+1/\kappa},\\
T  & = & T_{0}\left(  \frac{t_{0}}{\tau}\right)  ^{1/\kappa}\frac
{1}{\mathcal{V}(s)},\label{e:tsol1}%
\end{eqnarray}
where $p_{0}=n_{0}T_{0}$.
Thus we have generated a new family of exact solutions of
relativistic hydrodynamics: a new hydrodynamical solution is assigned to each
non-negative function $\mathcal{V}(s)$. It can be checked that the above
solutions are valid also for massive particles, the form of the solution is
independent of the value of the mass $m$. The form of solutions depends
parametrically on $\kappa$, that characterizes the equation of state.

\subsection{Analysis of the solutions}

The pressure and the flow profiles 
of the above 1+1 dimensional relativistic hydro solution
are the same as in the boost-invariant solution. In the case of 
$\mathcal{V}(s)=1,$ we recover the Hwa-Bjorken 
boost-invariant solution of refs.~\cite{scale-invariant,Bjorken}. 
In this limiting case, the pressure, 
the density and the temperature profiles depend only 
on the longitudinal proper time $\tau$.

In the general case, our solution contains a characteristic scale defining
parameter in the longitudinal direction, $\dot{Z}_{0}$, and an arbitrary
scaling function $\mathcal{V}(s)$. Thus we have an infinitely rich new family
of solutions. Let us try to determine the physical meaning of the scaling
function $\mathcal{V}(s)$.

In order to do this we evaluate the single particle spectra corresponding to
the new solutions. Here we neglect any possible dynamics in the transverse
directions, as usual in case of applications of the boost-invariant solution. 
The four-velocity field of our solutions thus becomes $u^{\mu}=(\cosh
\eta,0,0,\sinh\eta)$. 
The four-momentum of the observed particles with mass
$m$ is denoted by $k^{\mu}=(m_{t}\cosh$ $y,k_{x},k_{y},m_{t}\sinh$ $y)$. \
Let us assume
that particles freeze out at a constant longitudinal proper-time $\tau_{f}$,
for the sake of simplicity. This implies freeze-out at a constant pressure,
but at a space-time rapidity dependent temperature and density, and makes it
possible to continue the calculation analytically. The source function of
locally thermalized relativistically flowing particles in a Boltzmann
approximation can be written as
\begin{equation}
S(x,\mathbf{k}) =C({\eta})\, m_{t}\cosh(\eta-y) \, 
	n(x)\, \exp\left( -k^{\mu}u_{\mu }/T\right)  \, \delta(\tau-\tau_{f}),
\end{equation}
where $C(\eta)$ is an $\eta$ dependent normalization factor, 
given by the condition that 
$\int d\mathbf{k}/E\, S(x,\mathbf{k})=n(x)\delta(\tau-\tau_{f})$, 
which implies that
\be
C(\eta) = \left\{4\pi m^2 T(\tau_f,\eta) K_2[m/T(\tau_f,\eta)]\right\}^{-1}, 
\ee
where 
$ K_\nu(z) = \int_0^\infty dz \exp(-z \cosh t) \cosh(\nu t)$
is the modified Bessel function of the second kind.
\bigskip

The single particle spectrum can be calculated from the emission function as
\begin{equation}
E\frac{d^{3}N}{d\mathbf{k}}=\int\tau d\tau d\eta S(x,\mathbf{k}).
\end{equation}

Substituting our family of new solutions, 
and using $\mathcal{T}(x)=1/\mathcal{V}(x)$, 
we obtain
\begin{eqnarray}
S(x,\mathbf{k})  & = & \, C(\eta) \, m_{t}\cosh(\eta-y) \, n(x) 
f_B(x,\mathbf{k}) \\
f_B(x,\mathbf{k}) & = & 
\exp\left[  {-\frac{m_{t}\cosh(\eta-y)}{T_{0}}}\left(  {\frac{\tau}{t_{0}}
}\right)^{1/\kappa}{\mathcal{V}(\frac{\tanh^{2}\eta}{\dot{Z}_{0}^{2}}
)}\right]  \delta(\tau-\tau_{f}).
\end{eqnarray}
We are interested in the coupling between the measurable rapidity distribution
and the rapidity dependence of the effective temperature in the transverse
directions as obtained from our new family of solutions. 
We assume that $\mathcal{V}(s)$ is a slowly varying function, i.e. 
$d\log\mathcal{V}(s)/ds\ll1$ in the region of interest. This assumption 
implies that the point of maximal emissivity is located at $\overline{\eta}=y$ 
with correction terms of $\mathcal{O}(d\log\mathcal{V}(s)/ds)$ 
The measurable single-particle spectra can be written as
\begin{eqnarray} 
E\frac{d^{3}N}{d\mathbf{k}}  & = & 2 C(y) \, n_{0}t_{0}\, \mathcal{V}\left(
\frac{\tanh^{2}y}{\dot{Z}_{0}^{2}}\right)  
K_1[m_{t}/T_{\mathrm{eff}}\left(  y\right) ],\\
\frac{dN}{dy}  & = & n_{0}t_{0} \mathcal{V}\left(
\frac{\tanh^{2}y}{\dot{Z}_{0}^{2}}\right).
\label{e:dndy}
\end{eqnarray}
where\\
\begin{equation}
T_{\mathrm{eff}}(y)=\frac{1}{\mathcal{V}\left(  \frac{\tanh^{2}y}{\dot{Z_0}^{2}%
}\right)  }T_{0}\left(  \frac{t_{0}}{\tau_{f}}\right)  ^{1/\kappa}.
\label{e:teffy}
\end{equation}

Note that the $\mathcal{V}$ function is a free fit function that describes the
measurable rapidity distribution, including characteristic scales of the size
of $\dot{Z}_{0}$. 

We see that the slope parameter for transverse mass distribution
$T_{\mathrm{eff}}$ is related to the rapidity distribution as%
\begin{equation}
T_{\mathrm{eff}}(y)=T_{0}\left(  \frac{t_{0}}{\tau_{f}}\right)  ^{1/\kappa
}{\frac{dN/dy\left(y=0\right)}{dN/dy}}.
\end{equation}
Figures  1 and 2 illustrate the calculated behavior of the effective temperature
distribution as a function of rapidity for a single Gaussian-like  
and  a double
Gaussian-like ansatz for the measurable rapidity distribution. 

\begin{figure}[tbp]
\vspace{9.5cm}
\includegraphics{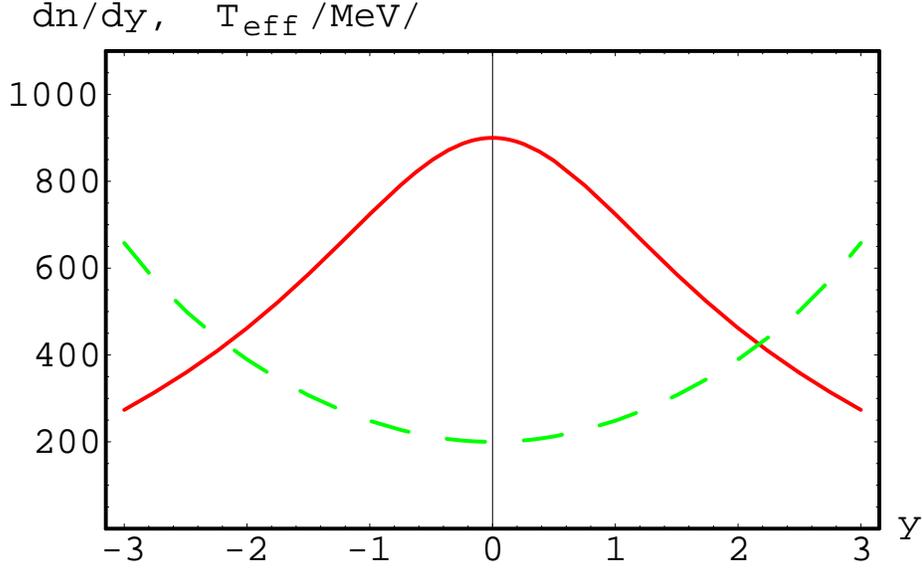}
\vspace{-1.3cm}
\caption{
Rapidity distribution $dN/dy$ and effective temperature distribution
$T_{\rm eff}(y)$ as a function of rapidity $y$, as obtained  from
a new family of solutions of (1+1) dimensional relativistic
hydrodynamics. Here we use the scaling function
$\NN(s) = (1 - s)^{(1/4)}$, using a scale parameter $\dot{Z_0} =\tanh(4)$,
$n_0 t_0 = 900$ and $T_0 (t_0/\tau_f)^{1/\kappa} = 200$ MeV, corresponding 
to a single maximum in
the rapidity distribution $dN/dy$. 
The analytic expressions are given by eqs.~(\ref{e:nsol},
\ref{e:tsol},\ref{e:dndy},\ref{e:teffy}).}
\end{figure}

An interesting aspect of this new 1+1 dimensional 
solution is that the shapes of the rapidity
distribution $dN/dy$ and temperature distribution are coupled:
the larger the rapidity density, the smaller the effective temperature.
Choosing the effective temperature distribution $T_{\mathrm{eff}}(y)$ to be
flat, we recover the Hwa-Bjorken 1+1 dimensional solution,
and the $dN/dy$ rapidity distribution also becomes flat, rapidity independent. 
This behavior is expected to appear in high energy heavy 
ion collisions in the infinite bombarding energy limit.

\begin{figure}[tbp]
\vspace{9.5cm}
\includegraphics{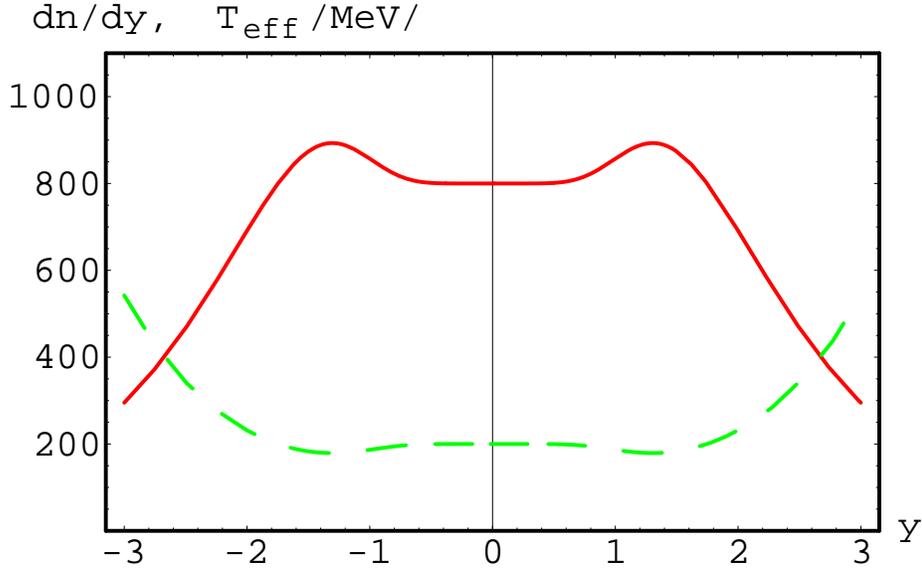}
\vspace{-1.3cm}
\caption{Same as Fig. 1 but utilizing a different  form of the
scaling function, 
$\NN(s) = \sqrt{1 + 1.6  s^4 - 2.6 s^8}$, using a scale parameter 
$\dot Z_0^2 = 1$, $n_0 t_0 = 800$  and $T_0 (t_0/\tau_f)^{1/\kappa} = 200$ 
MeV, corresponding to a two-peaked rapidity distribution.}
\end{figure}

\section{Cylindrically symmetric solutions \label{s:cyl}}

In this section, we describe a new family of exact analytic solutions of 
relativistic hydrodynamics, with cylindrically symmetric flow, 
overcoming the second of shortcoming of the well known boost-invariant 
Hwa-Bjorken solution~\cite{scale-invariant,Bjorken}.
However, we do not address both 
shortcomings simultaneously yet. 
The physical motivation for this study is to consider the time evolution of 
central collisions in ultra-relativistic heavy-ion physics within the 
framework of an analytic approach. 
From now on, $x^\mu = (t, r_x, r_y, r_z)\equiv (t,{\bf r})$ and 
$k^\mu = (E, k_x, k_y, k_z) \equiv (E,{\bf k})$ with $E^2 - {\bf k}^2 = m^2$.

As we are primarily interested in the effects of finite transverse 
size and the development of transverse flow, we assume that the 
longitudinal flow component is boost-invariant, 
\begin{equation} 
v_z(t,r_z)  =  \frac{r_z}{t} \label{e:vz}.
\end{equation} 
We search for self-similar solutions, that are scale dependent in 
the transverse directions, and depend only on the transverse radius 
variable $r_t = \sqrt{r_x^2 + r_y^2}$ through the scaling variable  
\begin{equation} 
	s=\frac{r_x^2 + r_y^2}{R^2}, 
\end{equation} 
and the longitudinal proper time $\t = \sqrt{t^2 - r_z^2}$ and assume that, in the frame where $v_z=0$ (longitudinal proper frame), the transverse motion corresponds to a Hubble type of self-similar transverse expansion, 
\bea
v_x^*(\t,r_z)& = & {\frac{\dot{R}(\t)}{R(\t)}}r_x\,, \qquad 
v_y^*(\t,r_z) \, = \, {\frac{\dot{R}(\t)}{R(\t)}}r_y\,, \label{e:vxy} 
\eea
where $\dot R = dR(\t)/d\t$ and hereafter we will designate by 
starred symbols the variables in the longitudinal proper frame. We 
assume that the scale $R$ depends on time only through the 
longitudinal proper time, $\t$. 

In a relativistic notation, the above form may be parametrized as  
\bea
u^\mu & = & (\cosh\zeta \cosh\xi, 
		\sinh\xi \frac{r_x}{r_t}, \sinh\xi \frac{r_y}{r_t},
		\sinh\zeta \cosh\xi),\\ 
\tanh\xi& = &{\frac{\dot{R}(\t)}{R(\t)}}r_t=v_t^*=\gamma_lv_t\,,\qquad \mbox{\rm or}\quad 
\cosh\xi\, = \,\frac{1}{\sqrt{1 - \dot R^2 s}} \equiv \gamma_t^*\,,\\ 
\cosh\zeta & = & \frac{t}{\t} \equiv \gamma_l\,.
\eea
The space-time rapidity $\eta$ is still defined by eq.~(\ref{e:eta1}). 
For a scaling longitudinal flow we obtain 
$\zeta = \eta\,$. 
Using the above ansatz for the flow velocity distribution, 
we find that the continuity equation is solved by the form
\begin{equation} 
n(t,r_x,r_y,r_z) = n_0 
	\left(\frac{\tau_{z0}R_0^2}{\t R^2}\right)
	\frac{1}{\cosh\xi} \N(s), \label{s:cont}
\end{equation} 
where $\N(s)$ is an arbitrary non-negative function of the scaling 
variable $s$ and $n_0\,$, $\tau_{z0}$ and $R_0$ are normalization 
constants. We use the convention $n_0 = n(t_0, 0,0,0)$, 
$\tau_{z0}=\tau_z(t_0,r_{z0})$ and $R_0 = R(\tau_{z0})$, where 
$r_{z0}$ is such that, together with $t_0\,$, satisfies 
eq.~(\ref{e:vz}). This implies that $\N(s=0) = 1$. The temperature 
equation, eq. (\ref{e:rT}), is solved by 
\begin{equation} 
T(t,r_x,r_y,r_z) = T_0 
		\left(\frac{\tau_{z0} R_0^2}{\t R^2}\frac{1}{\cosh\xi}\right)^{1/\kappa} 
	\T (s).
\end{equation} 
The constants of normalization are 
$T_0 = T(t_0,0,0,0)$ and $\T (0)= 1$. 
We find that the solution is independent
of the form of the function $\T (s)$, provided that $\T (s) > 0$.

Using a similar technique as in section 3, we obtain a transcendental equation 
for $\dot R^2$, and $s$.  This equation has a particular solution if 
\begin{equation} 
	\dot R = \dot R_0 = \mbox{\rm Const} . \label{rdot}
\end{equation} 
In this case, the acceleration of the radius parameter vanishes, 
$\ddot R = 0$, and the solution is 
$R = R_0 + \dot R_0 (\t - \tau_{z0})$. 
The relativistic Euler equation reduces to 
\begin{equation} 
 \left(1 + \frac{1}{\kappa} \right) 
  \left(\frac{R \dot R}{\tau_z} + 3\dot R^2\right) 
	 = 2(1 - s \dot R^2)  \left[\log\N(s) \T(s)\right]^\prime\ , 
 \label{euler}
\end{equation} 
where the lhs depends only on $\t$ while the rhs is only a function
of the variable $s$, hence both sides are constant.
This implies that
$
	\frac{R}{\t} = \dot R_0\, , 
$
thus $R_0 = \dot R_0 \tau_{z0}\,$. Thus the origin of the time axis 
(fixed by the assumption of the scaling longitudinal flow 
profile) coincides with the vanishing value of the transverse radius 
parameters. 

The solutions can be casted in a relatively simple form by introducing
the proper time $\tau$,  
\bea
\tau & =&  \sqrt{\t^2 - r_t^2} \, = \, 
	\sqrt{t^2 - r_x^2 - r_y^2 - r_z^2}. \label{e:tau}
\eea
Using this natural variable we find that
\bea
	{\mathbf v}  & = & \frac{{\mathbf r}}{t} , \label{e:vsol}\qquad \mbox{\rm or} \qquad 
	 u^\mu \,  = \,  \frac{x^\mu}{\tau}.
\eea
Thus the velocity field of our solution corresponds to the flow
field of the spherically symmetric scaling solution and to the Hubble
flow of the Universe. However,
in the scaling solution the temperature and the pressure 
distributions are dependent only on the proper time $\tau$, while 
in our case both the density and the temperature distributions are 
generally dependent on the scale variable $s$ in the  transverse 
direction. 

As the solution is relativistic, and it is defined in the positive light-cone, given by $\tau \ge 0$, we obtain a constraint for the 
transverse coordinate, $r_t \le \t$. This together with 
the solution for the scale $R$, implies that 
the scaling variable has to satisfy the constraint 
$s \dot R_0^2 \le 1$, which corresponds to the limitation that the 
velocity of the fluid can not exceed the speed of light. 

By substituting $R = \dot{R}_0 \t $ into the Euler equation, eq.(\ref{euler}), one obtains 
\begin{equation} 
\frac{d}{ds}
 \log\left[(1-s\dot R_0^2)^{2(1+1/\kappa)}\N(s)\T(s)\right]=0\,, 
\end{equation} 
which gives, together with the condition $\N(0)\T(0) = 1$,
\begin{equation} 
\N(s) \T(s) = (1 - \dot R_0^2 s)^{-  2(1 + 1/\kappa)}. \label{e:nt}
\end{equation} 
In this family of solutions, the scaling functions for the 
temperature and the density distribution are thus not independent. 
However, a constraint is given for their product, hence one of them
can be chosen as an arbitrary positive function.
For clarity, let us introduce new forms of the scaling functions as 
\bea
	\TT (s) & =& \T (s)  (1 - \dot R_0^2 s)^{2/\kappa}, \\
	\NN (s) & = & \N (s)  (1 - \dot R_0^2 s)^2.
\eea
Then the constraint can be casted to the simple form of
$\NN(s) \TT(s) = 1$.
This construction for the scaling functions of the transverse density
and temperature profiles coincides with the method, that we developed
for the solution of the relativisitic hydrodynamical equations in
the (1+1) dimensional problem, but here the transverse flow 
has a two-dimensional distribution, so the exponents and the scaling
variables had to be re-defined accordingly.

Let us summarize our new family of solutions of the 1+3 dimensional
relativistic hydrodynamics for cylindrically symmetric systems
by substituting the results to the density,
temperature and pressure profiles. 

We obtain
\bea
{\mathbf v} & = & \frac{\mathbf r}{t} , 
		\quad \mbox{\rm for} \quad |{\bf r}| \le t, \\
s & = & \frac{r_t^2}{\dot R^2_0 \t^2}, 
		\quad \mbox{\rm for} \quad {r_t} \le \t, \\
n(t,{\mathbf r}) & = & n_0 \left(\frac{\tau_{z0}}{\tau}\right)^3 \NN(s), \label{e:nsol} \\
p(t,{\mathbf r}) & = & p_0 \left(\frac{\tau_{z0}}{\tau}\right)^{3 + 3/\kappa}, \\
T(t, {\mathbf r}) & = & T_0 \left(\frac{\tau_{z0}}{\tau} \right)^{3/\kappa}\frac{1}{\NN (s)}, \label{e:tsol} 
\eea
where $p_0 = n_0 T_0$. 
Note that the scaling variable $s$ is invariant for boosts in the 
longitudinal direction, and it is rotation-invariant in the 
transverse direction, {\it but $s$ is not boost-invariant in the 
transverse directions}. Hence we have generated  cylindrically 
symmetric, longitudinally boost invariant solutions of relativistic 
hydrodynamics. In the longitudinal direction, these solutions are 
homogeneous, boost-invariant and also scale-invariant. Due to this 
reason, the observable rapidity distribution is 
\begin{equation} 
	\frac{dN}{dy} = \mbox{\rm const},
\end{equation} 
a flat distribution, corresponding to the ultra-relativistic
nature of the solution in the longitudinal direction (where 
$y = 0.5\log[(E+k_z)/(E-k_z)]$ is the rapidity of a particle with 
four-momentum $(E,{\bf k}) $ and $dn/dy$ is the rapidity 
distribution of particle density).

A new hydrodynamical solution is assigned to each non-negative 
function $\NN (s)$, similarly to the cases of the non-relativistic 
solutions of ref.~\cite{cssol} and the 1+1 dimensional relativistic 
solution of the previous section. Note that the solutions are 
valid also for massive particles, the form of the solution is 
independent of the value of the mass $m$. The form of solutions 
depends parameterically on $\kappa$, that characterizes the 
equation of state.

We have obtained new solutions of the (1+3) dimensional relativistic 
hydrodynamical equations which describe a self-similar, streaming 
flow. In the case of $\dot R = 1$ and $\NN (s) = 1$ we recover 
the spherically symmetric scale-invariant solution. This means that, 
in this limiting case, the pressure, the density and the temperature 
profiles depend only on the proper time $\tau$. 
In general case, however, our solution depends not only on the characteristic scale $R$ 
 but also on the arbitrary scaling function $\NN (s)$.

\section{Summary}
We have found a new family of both 1+1 dimensional, longitudinally 
expanding, and  1+3 dimensional, cylindrically symmetric, adiabatic solutions of 
relativistic hydrodynamics with conserved particle number. These 
families of solutions solves the continuity equation and the 
conservation of the energy - momentum tensor of a perfect fluid, 
assuming simple equations of state, given by Eqs.(\ref{e:eos1}) and 
(\ref{e:eos2}). The mass of the particles $m$ and 
$\kappa=\partial\ep/\partial p = 1/c_s^2$ are free parameters of the solution.
The well-known scale-invariant solution, has been obtained in the $m=0$ 
approximation. Interestingly, our generalizations resulted in 
{\it additional freedom} in the solution.

In the new 1+1 dimensional hydro solutions, the flow field coincides 
with that of the Hwa-Bjorken solution. In principle, the shape of the 
measurable rapidity distribution, $dN/dy$ plays the role of an arbitrary 
scaling function in our solution, and we obtain that the effective 
temperature of the transverse momentum distribution becomes rapidity 
dependent. Assuming that $dN/dy$ is a slowly varying function of the 
rapidity $y$, we find that the effective temperature is proportional 
to the inverse of the rapidity distribution, 
$T_{\mathrm{eff}}(y)\propto(dN/dy)^{-1}$.

In 1+3 dimensions, even the flow velocity field deviates from Hwa-Bjorken 
solution. We find that the {\it only} exact solution in the considered 
class corresponds to a scaling 3-dimensional flow, similar to the Hubble 
flow of the Universe. 
Although the pressure distribution is only proper-time dependent,
this pressure is a product of the local number density 
and the local temperature, 
hence one of these can be chosen in an arbitrary manner.

The essential result of our paper is that we found a rich family of 
exact analytic solutions of relativistic hydrodynamics that 
contain both a longitudinal Hwa-Bjorken flow (that is frequently utilized
in estimations of observables in high energy heavy ion collisions)
and a relativistic transverse flow (whose existence is evident
from the analysis of the single particle spectra at RHIC and SPS 
energies~\cite{cracow,cracow-rev,cracow-school,bl-bang}).


{\it Acknowledgments:}  
T. Cs. would like to thank L.P. Csernai, B. Luk\'acs and 
J. Zim\'anyi for inspiring discussions during the initial phase of 
this work, and to Y. Hama, G. Krein and S. S. Padula for their kind 
hospitality during his stay at USP and IFT, S\~ao Paulo, Brazil. 
This work has been supported by a NATO Science Fellowship (T. Cs.), 
by the OTKA grants T026435, T029158, T034269 and T038406 of Hungary, 
the NWO - OTKA  grant N 25487 of The Netherlands and Hungary, and 
the grants FAPESP 00/04422-7, 99/09113-3, 02/11344-8, 
PRONEX 41.96.0886.00, FAPERJ E-26/150.942/99, and CNPq, Brazil. 


\end{document}